\documentclass[pre,amsmath,amssymb,floatfix,twocolumn,superscriptaddress,nofootinbib]{revtex4}
\usepackage{graphicx}
\usepackage{epsfig}
\usepackage{dcolumn}
\usepackage{bm}

\def\(({\left(}
\def\)){\right)}
\def\[[{\left[}
\def\]]{\right]}

\newcommand{\beq}{\begin{equation}}
\newcommand{\eeq}{\end{equation}}
\newcommand{\barr}{\begin{eqnarray}}
\newcommand{\earr}{\end{eqnarray}}
\newcommand{\bei}{\begin{itemize}}
\newcommand{\eei}{\end{itemize}}

\newcommand{\s}{\sigma}

\newcommand{\ra}{\rangle}
\newcommand{\la}{\langle}

\renewcommand{\a}{\alpha}

\def\us{\underline{\sigma}}

\begin{document}

\title{On the relation between kinetically constrained models of glass dynamics \\
and the random first-order transition theory}
\author{Laura Foini} 
\affiliation{Universit\'e Pierre et Marie Curie -- Paris VI, 
LPTHE, Tour 13 5\`eme \'etage, 4 Place Jussieu, 75252 Paris Cedex 05, France} 
\author{Florent Krzakala}
\affiliation{ESPCI ParisTech, CNRS UMR 7083
  Gulliver, 10 rue Vauquelin, 75005 Paris, France}
\author{Francesco Zamponi}
\affiliation{LPT,
\'Ecole Normale Sup\'erieure, CNRS UMR 8549, 24 Rue Lhomond, 75005 France}

\begin{abstract}
In this paper we revisit and extend the mapping between two apparently different classes
of models.
The first class contains the
prototypical models described --at the mean-field level--
by the Random First Order Transition (RFOT) 
theory of the glass transition, 
called either ``random XORSAT problem'' (in the information theory community) 
or ``diluted $p$-spin model'' (in the spin glass
community), undergoing a single-spin flip Glauber dynamics.  
The models in the second class are Kinetically Constrained Models (KCM): 
their Hamiltonian is that of independent spins in a constant magnetic field,
hence their thermodynamics is completely trivial,
but the dynamics is such that only groups of spin can flip together, 
thus implementing a kinetic constraint that induces a non-trivial dynamical 
behavior.
A mapping between some representatives of these two classes has been known for long.
Here we formally prove this mapping at the level of the master equation, and we
apply it to the particular case of Bethe lattice models. This allows us to show that
a RFOT model can be mapped exactly into a KCM.
However, the natural order parameter for the RFOT model, 
namely the spin overlap, is mapped into a very complicated non-local function in
the KCM. Therefore, if one were to study the KCM without knowing of 
the mapping onto the RFOT model, one would guess that its
physics is quite different from the RFOT one. 
Our results instead suggest that these two apparently different 
descriptions of the glass
transition are, at least in some case, closely related.
\end{abstract}

\maketitle

\section{Introduction}

The Random First-Order Transition (RFOT)~\cite{KTW89,CC05,LW07,BB09}
and the dynamic facilitation theory based on 
Kinetically Constrained Models (KCM)~\cite{Ga02,RS03,GST10} are often viewed 
as alternative pictures of glass formation~\cite{RS03,Ca09,BB11,WG04}.
Indeed, in the former the glass transition is induced by an underlying thermodynamic
transition, the so-called ``Kauzmann transition'', while in the latter the thermodynamics
is assumed to be completely trivial, and the glass transition is a purely dynamical
arrest induced by complex kinetic constraints\footnote{
Note that sometimes (especially in the mathematical literature) a more stringent definition
of KCM is employed. Namely, one requires that the dynamics of a finite size system
has absorbing (or ``blocked'') states. In the following, we will just define a KCM as 
a system that has a trivial thermodynamics, and a complex dynamic behavior induced by
specific dynamic rules.
}.

The RFOT theory\footnote{
In this paper, by RFOT theory we denote the ensemble of theories that can be related
to the exact solution of mean-field spin glass models (therefore including mode-coupling
theory, molecular liquid replica theory, and finite-dimensional extensions). 
In particular, we will mostly be
concerned with mode-coupling-like dynamics, but we will keep calling it RFOT dynamics.
} is based on the analysis of a particular class
of mean-field spin glass models with $p$-spin interactions~\cite{KTW89}.
It has been known for long~\cite{Ga02,JBG05,JG05} that some particular finite-dimensional 
version of these $p$-spin models (the so-called ``plaquette models'') can be explicitly mapped 
onto KCMs. However, the physics of these models, at least in two dimensions, 
seems to be quite different from
the RFOT one~\cite{Ga02,JBG05,JG05,CB12}, thereby preventing one to establish a direct mapping
between the two approaches.

The aim of this work is to discuss an example in which, on the contrary, a
KCM can be exactly mapped into a model
described by the RFOT theory. In order to do this, we first generalize the mapping
discussed in~\cite{Ga02,JBG05,JG05,CB12} to an arbitrary geometry, with the only requirement
that {\it i)} each spin is involved in exactly $p$ interactions, {\it ii)} each interaction involves 
exactly $p$ spins, and {\it iii)} the ground state is unique. By using the method developed by 
Gosset~\cite{Gosset} for quantum Hamiltonians,
we show that the Fokker-Plank operator of the $p$-spin model can be mapped into the one of the KCM.

We then focus on a Bethe lattice geometry, because this case allows for a very complete and detailed analysis,
and in particular the Bethe lattice $p$-spin model is exactly described by the RFOT theory.
More precisely, we consider the two following models:
\begin{itemize}
\item
The first model is a spin model defined on a random regular 
factor graph (or Bethe lattice) geometry, where each spin is randomly connected to three
interactions and each interaction involves three spins.
The dynamics is a single spin-flip Glauber dynamics.
The model is known as ``random 3-XORSAT'' in the information theory community,
and as the diluted $3$-spin in the spin glass community. 
It was studied intensively by both communities~\cite{FMRWZ01,FLRZ01,xor_1,xor_2,xor_replica},
and it was shown that the dynamical spin-correlations show a plateau in time 
that is naturally interpreted from a thermodynamical point
of view in terms of the emergence of disconnected ergodic 
states that trap the system's dynamics (see Fig.~\ref{spin-overlap}).
Rigorous results have in fact shown that for such a system the Gibbs-Boltzmann measure
can be decomposed into disconnected components and purely dynamical
quantities are recovered from a thermodynamical study~\cite{xor_1,xor_2,MoSe}.
We will refer to this model simply as ``XORSAT'' in the following.
\item
The second model is defined on the same geometry as the first one.
Its Hamiltonian is given by non-interacting
spins in a magnetic field, and its dynamics only allows a simultaneous flipping
of a group of three spins connected to the same ``interaction''. 
This model displays the typical behavior of a Bethe lattice KCM~\cite{SBT05},
i.e. it has trivial equilibrium thermodynamical
properties but the dynamic rules lead to a complex dynamical behavior characterized by a finite-temperature
dynamical transition.
As in other KCMs, despite the simplicity of
the equilibrium Gibbs-Boltzmann measure, the dynamical correlations display slow relaxation
and characteristics that resemble those of glassy systems.
In the following we will refer to this second model 
as the KCM.
\end{itemize}
Thanks to the general mapping discussed above, we show that these two models are in fact completely equivalent:
the master equations describing their dynamics can be mapped one into the other by a suitable
transformation, that maps the groups of three interacting spins (``plaquettes'') 
of XORSAT into the spins of the KCM.
We show that while different observables
share a similar dynamical behavior, and in particular a plateau in their time correlations associated with the dynamical
transition, they crucially differ once that they are chosen as order parameters 
in a static computation.

Therefore, some observables fail to identify the ergodicity breaking.
Remarkably, this is the case for all local observables in the KCM: detecting ergodicity breaking in the KCM therefore requires
the use of non-local (hence very unnatural and complicated) observables.
To illustrate this point, we will consider two distinct order parameters for the XORSAT model: the spin overlap $q_s$ and the
plaquette overlap $q_p$. We will show that $q_s$ is a good order parameter, while $q_p$ is not. 
The plaquette overlap for XORSAT becomes the spin overlap for the KCM, therefore the spin overlap is not a good order parameter
for the KCM. Moreover, the spin overlap for XORSAT becomes a very complicated non-local function for the KCM; therefore,
detecting the glass transition by means of a static computation in the KCM requires the use of a very unintuitive order parameter.
If one were to study the KCM without knowing of the underlying mapping on XORSAT, it would be impossible to identify the correct
order parameter, and one would conclude that the dynamic transition cannot be detected via a static computation, and that the dynamical
facilitation picture is completely different from the RFOT one. 

Our results suggest that this conclusion might be wrong: at least for the model we investigated, the two scenarios are dual to each other.
The same could be true in more general models~\cite{KPS97}.

The paper is organized as follows. In section~\ref{sec:Gmapping} we establish in general the mapping between the 
Fokker-Planck equations describing the two classes of models. In section~\ref{sec:Bethe} we focus on the Bethe lattice
models and present the results for both dynamic correlation functions and the thermodynamic RFOT potential.
Finally, in section~\ref{sec:discussion} we discuss our results and their implications.

\section{Mapping $p$-spin models into KCMs}
\label{sec:Gmapping}

In this section we show how to characterize the equilibrium dynamics
of the models and how to construct the mapping between the two.

Let us first outline the general Fokker-Planck equation for the stochastic 
dynamics of $N$ Ising spins, $\sigma_i=\pm 1$ and $i=1,\dots,N$, 
described by a generic energy function $E(\us)$.
We define $P(\us,t)$ the probability that at time $t$ the system is in the configuration $\us$. 
Suppose that the elementary dynamics is induced by the flipping of a set of variables 
$k = \{ i^1_k, i^2_k, \cdots, i_k^{|k|}\}$
of size $|k|$,
with rate $w(\us^{(k)},\us)$, where $\us^{(k)}$ represents the configuration
that is obtained from $\us$ by the flip of the set $k$.
Then, the probability $P(\us,t)$ satisfies the equation:
\beq\label{FP-eq}
\begin{split}
\frac{\partial P(\us,t)}{\partial t} & = - \sum_k w(\us^{(k)},\us) P(\us,t) \\
& + \sum_k w(\us,\us^{(k)}) P(\us^{(k)},t) ,
\end{split}\eeq
where the sum over $k$ stems for all possible sets of variables.
In order to reach equilibrium the rates have to satisfy the detailed balance condition:  
\beq\label{DBC}
w(\us^{(k)},\us) e^{- \beta E(\us)} = w(\us,\us^{(k)})  e^{- \beta E(\us^{(k)})} \ ,
\eeq
and in the following we will adopt the common choice 
\beq\label{our_choice}
w(\us^{(k)},\us) = w_0 e^{- \frac{\beta}{2} (E(\us^{(k)}) - E(\us))} \ ,
\eeq
with $w_0 =1$ in order to set the unit of time.

Defining the matrix $H$ such that $\langle \us | H | \us' \rangle 
=  \sum_k w(\us^{(k)},\us) \delta(\us,\us') -  \sum_k w(\us,\us^{(k)})  \delta(\us^{(k)},\us')$,
and $| P(t) \rangle$ the probability vector such that $\langle \us |  P(t) \rangle =  P(\us,t)$,
Eq.~(\ref{FP-eq}) can be written as a Schr\"odinger equation in imaginary time:
\beq\label{EqFP}
\frac{\partial | P(t) \rangle}{\partial t} = - H | P(t) \rangle \ .
\eeq 
It is often convenient to introduce the transformation induced by the diagonal matrix $U$, such that 
$\langle \us | U | \us' \rangle =  e^{\frac{\beta}{2} E (\us)} \delta(\us,\us')$, which turns
Eq.~(\ref{EqFP}) into:
\beq\label{EqFPh}
\frac{\partial | \tilde{P}(t) \rangle}{\partial t} = - \hat{H} | \tilde{P}(t) \rangle \ ,
\eeq
with $ | \tilde{P}(t) \rangle = U | P(t) \rangle$ and $ \hat{H} = U H U^{-1}$. The advantage of this formulation is
that the matrix $\hat{H}$ is Hermitian.

Under the condition that the rates satisfy Eq.~(\ref{DBC}), 
one can verify that the equilibrium Gibbs-Boltzmann measure 
$P(\us,t) = P_{GB}(\us) \propto \exp(- \beta E(\us))$ is
a stationary solution of Eq.~(\ref{FP-eq}) and it represents the ground state
of the Fokker-Planck operator $H$ in Eq.~(\ref{EqFP}), and the same holds
for $\tilde{P}_{GB}(\us) \propto \exp(- \beta E(\us)/2)$ and $\hat{H}$.

\subsection{XORSAT model with spin-flip dynamics}

We consider a model of $N$ Ising spins characterized by $3$-body interactions
with Hamiltonian:
\beq\label{Hpspin}
E_{XOR}(\us) = - \sum_{a=1}^{M} \sigma_{i_a^1} \sigma_{i_a^2} \sigma_{i_a^3}
\eeq
where $a$ labels the interactions, each one involving a triplet of spins.
Such Hamiltonian
represents a particular case ($p=3$) of a general class of spin models 
with $p$-body interactions that in computer science is also known as XORSAT
problem.

The graphical representation of the problem can be obtained through the introduction of a 
``factor graph'', by assigning a
``variable node'' to each spin $i$ and an ``interaction node'' to each interaction $a$ and
 connecting every spin to the interaction nodes in which it takes part. Here we assume that each
 spin is connected to exactly three interactions, hence the connectivity of spin variables is $c=3$.
 For the Hamiltonian in Eq.~(\ref{Hpspin}), the connectivity of each interaction node is $p=3$ and
the number $M$ of these nodes must satisfy $M p = N c$, then $M=N$.
In particular,
from Eq.~(\ref{Hpspin}), the ferromagnetic configuration $\sigma_i=1$ $\forall i=1,\dots,N$ 
minimizes all the interactions $a=1,\dots,M$ 
and in what follows we will restrict to the case where this is the only ground state
(we will come back to this point in section~\ref{sec:Bethe}).

For this model we study the dynamics induced by single spin flips, 
so the sum over $k$ in Eq.~(\ref{FP-eq}) is simply the sum over all the spins and
the rates $w(\us^{(i)},\us)$ connect configurations differing only for a single spin $\sigma_i$. 
Then, according to Eqs.~(\ref{our_choice}) and (\ref{Hpspin}), 
$w_{XOR}(\us^{(i)},\us)  = e^{- \beta \sum_{a \in \partial i}  
\sigma_{i_a^1} \sigma_{i_a^2}  \sigma_{i_a^3}}$, where 
here and in the following $a \in \partial i$ indicates the neighborhood of $i$, i.e. the interactions $a$
to which the site $i$ is connected.
Performing the operations outlined in the previous section, one arrives to the following form for the
Hermitian Fokker-Planck operator that appears in Eq.~(\ref{EqFPh}):
\beq\label{HFP-pspin}
\hat{H}_{XOR} = \sum_{i=1}^N e^{ - \beta \sum_{a \in \partial i}  
\hat{\sigma}_{i_a^1}^z \hat{\sigma}_{i_a^2}^z \hat{\sigma}_{i_a^3}^z} - \sum_{i=1}^N \hat{\sigma}^x_i \ ,
\eeq
where $\hat{\sigma}^{x,y,z}_i$ are the standard Pauli matrices.

\subsection{Kinetically constrained model}\label{Sec:KCM}

We now introduce the kinetically constrained model (KCM), using different
notations that will be suitable for the mapping to be discussed in the next section.
We consider a system of $N$ Ising spins $s_a = \pm 1$ 
whose energy function reads:
\beq\label{energy-h}
E_{KCM}(\underline{s}) = - \sum_{a=1}^{N} s_{a} \ .
\eeq
Due to the non-interacting nature of $E_{KCM}(\underline{s})$ the thermodynamical
properties of the system are independent of the geometry where
the system is defined. 

In order to specify the constrained dynamics,
we assume that the model is defined on a regular random factor graph, with 
$p=c=3$, exactly as before, and we define the dynamics by the flip of triplets of spins
$\{a_i^1, a_i^2, a_i^3 \}$ 
surrounding the same ``interaction'' node $i$ (hence we use a dual
notation with respect to the XORSAT model).
The set $k$ in the Fokker-Planck Eq.~(\ref{FP-eq}) is now the set of the three spin
identified by $i$ and, following our choice in Eq.~(\ref{our_choice}), 
the rates read $w_{KCM}(\underline{s}^{(i)},\underline{s}) = \exp(- \beta  \sum_{a \in \partial i} s_{a})$.
The Hermitian evolution operator that enters in the Fokker-Planck
Eq.~(\ref{EqFPh}) for the KCM can be expressed in terms of  
Pauli matrices as follows: 
\beq\label{HFP-field}
\hat{H}_{KCM} = \sum_{i=1}^N e^{ - \beta \sum_{a \in \partial i} \hat{s}_{a}^x } - \sum_{i=1}^N \hat{s}^z_{a^1_i}\hat{s}^z_{a^2_i}\hat{s}^z_{a^3_i} .
\eeq
Note that the choice of the indices $z$ and $x$ 
in Eq.~(\ref{HFP-field}) is just an arbitrary choice of basis for the representation of the spin
configurations and one can safely exchange the two indices.

\subsection{Mapping}
\label{sec:mapping}

Following the procedure outlined in~\cite{Gosset} in the context of quantum spin systems, 
in this section we describe the mapping between $\hat{H}_{XOR}$, defined on a
regular graph with $c=p$ and such that $E_{XOR}$ has a unique ground state, into $\hat{H}_{KCM}$
defined on the dual graph of the interaction nodes.

Consider the operator $\hat{H}_{XOR}$ defined in Eq.~(\ref{HFP-pspin}) for 
a given graph with a unique (ferromagnetic) ground state.
Thanks to this property, for each interaction $a$, it exists one and only one
configuration of spins $\us^a$ minimizing all the interactions but $a$.
To show this, let us introduce the $N\times N$ adjacency matrix
$A$ such that $A_{a i}=1$ if the
spin $\sigma_i$ enters in the interaction $a$ and $A_{a i}=0$ otherwise, and
define $\underline{y}^{a}$ such that $y^a_i = \frac{\sigma^a_i-1}{2}$. 
The matrix $A$ is invertible because by assumption the system 
$A \underline{y} = \underline{0}$ 
(that is equivalent to the condition of minimizing $E_{XOR}(\us)$) 
has a unique solution, corresponding to  $\underline{y}= \underline{0}$.
Defining the vector $\underline{e}^a$ such that $e^a_{b} = \delta_{a,b}$, 
one gets $\underline{y}^{a} = A^{-1} \underline{e}^a$.

Then, one can introduce the following operators, defined on the interaction nodes 
of the underlying graph of Eq.~(\ref{HFP-pspin}):
\beq\label{directmapping}
\begin{split}
& \hat{s}^x_a = \hat{\sigma}_{i_a^1}^z \hat{\sigma}_{i_a^2}^z \hat{\sigma}_{i_a^3}^z \ ,  \\
& \hat{s}^z_a = \prod_{i=1}^N (\hat{\sigma}^x_i)^{y_i^a} \ .
\end{split}\eeq
These are well defined spins operators
because they satisfy the correct commutation relations, 
$[\hat{s}^{x}_a,\hat{s}^{z}_{b}] = 0$ if $a \neq b$, $\{\hat{s}^{x}_a,\hat{s}^{z}_{a}\} = 0$
and $(\hat{s}^{x}_a)^2=(\hat{s}^{z}_a)^2=\mathbb{I}$,
where $\mathbb{I}$ is the $2\times 2$ identity matrix.
Moreover, it turns out that defining $\underline{e}^i$ such that $e^i_{j} = \delta_{i,j}$,
and using the relation
\beq
A \underline{e}^i = \sum_{a\in\partial i} \underline{e}^a \hspace{.3cm}\Rightarrow \hspace{.3cm}\
\underline{e}^i = A^{-1}\sum_{a\in\partial i} \underline{e}^a = \sum_{a\in\partial i} \underline{y}^a 
\eeq
the mapping above can be inverted as follows:
\beq\label{inversemapping}
\begin{split}
& \hat{\sigma}^x_i = \hat{s}^z_{a^1_i}\hat{s}^z_{a^2_i}\hat{s}^z_{a^3_i}  \ , \\
& \hat{\sigma}^z_i = \prod_{a=1}^N  (\hat{s}^x_a)^{y_i^a} \ .
\end{split}\eeq 
Then, enforcing the mapping in Eq.~(\ref{directmapping}) and (\ref{inversemapping}) 
in $\hat{H}_{XOR}$, one obtains the operator
 $\hat{H}_{KCM}$ of Eq.~(\ref{HFP-field}) defined on the dual lattice.
Note that here we discussed the mapping for $p=c=3$, but the extension to any
other value of $p=c$ is straightforward~\cite{Gosset}.

We have therefore shown that $\hat{H}_{XOR}$ defined on a given graph 
can be written as $\hat{H}_{KCM}$ defined on the dual graph by means of the transformation
(\ref{inversemapping}), and viceversa.
Note that this implies in particular that the ground state of
$\hat{H}_{XOR}$, 
\beq\label{GSXOR}
\langle \us | GS_{XOR} \rangle = \frac1{ \sqrt{Z} } e^{- \frac12 \beta E_{XOR}(\us)} \ ,
\eeq
whose square is the Gibbs distribution, can be written as 
the one of $\hat{H}_{KCM}$ on the dual graph:
\beq\label{GSKCM}
\langle \underline{s} | GS_{KCM} \rangle =  \frac{1}{\sqrt{Z}} e^{- \frac12 \beta E_{KCM}(\underline{s})}
\eeq
and moreover the partition functions of the two models (that correpond to the normalization of the ground states)
are identical:
\beq\label{Zmap}
Z = \sum_{\us} e^{- \beta E_{XOR}(\us)} = \sum_{\underline{s}}  e^{-\beta E_{KCM}(\underline{s})} \ .
\eeq

\section{Bethe lattice models}
\label{sec:Bethe}

Here we specialize our discussion to Bethe lattice models. 
Different definitions of Bethe lattice have been used in the literature,
and here we consider as a model of Bethe lattice a {\it random regular factor graph}:
we construct an ensemble of graphs 
by giving uniform probability to all the graphs characterized by $N$ spins and $M$ interactions
such that each spin appears exactly in $c=3$ interactions and each interaction connects 
precisely $p=3$ spins, and such that Eq.~(\ref{Hpspin}) has a unique ground state (the last condition 
is needed in order to use the mapping discussed in section~\ref{sec:mapping}).
We will be interested in the properties of the model in the thermodynamic limit $N=M\to\infty$.
In this limit, such graphs have the property of being locally
tree-like, as a Bethe lattice should: indeed, typical loops have length of order $\log N$. 
The lack of a finite dimensional geometry make
statistical models defined on such lattices intrinsincally mean-field like.
Most importantly this property allows for an analytical treatment via the so-called ``cavity method'',
and many results have been derived in this way~\cite{xor_1}.
The thermodynamical properties of Eq.~(\ref{Hpspin}) are self-averaging with respect 
to the ensemble of the graphs so defined and it turns out that in the case considered 
here, $k=c$, the ground state is non-degenerate with finite probability~\cite{JKSZ10}.
Hence, the restriction to graphs that have a unique ground state, that
has to be made here in order to derive the mapping,
does not alter the main properties of the model.
Note that the ensemble of regular random graphs with $c=p$ is self-dual,
hence the KCM will be defined on the same ensemble of graphs.
Note also that the two dimensional triangular lattice considered in \cite{Ga02}
represents a (very atypical) instance of such ensemble of graphs.

\subsection{XORSAT model}

The XORSAT model on the random regular factor graph with $p=c=3$ follows exactly
the RFOT scenario of the glass transition~\cite{xor_1,xor_2,ZM08}. It is characterized
by a dynamic transition at a temperature $T_d= 0.5098(2)$, below which ergodicity is broken and
the equilibrium dynamical correlation functions of local observables never relax.
In the RFOT scenario there is a second temperature $T_k$ at which a thermodynamic transition
to a replica-symmetry broken state takes place~\cite{KTW89}. However, here $T_k=0$, because of the
mapping onto independent spins that ensures that the free energy is analytic at all temperatures, see Eq.~(\ref{Zmap});
this can also be checked by an explicit replica or 
cavity computation~\cite{FMRWZ01,FLRZ01,xor_1,ZM08,KZ10a}.

\subsubsection{Spin correlations}

\begin{figure}
  \begin{center}
    \includegraphics[width=\columnwidth]{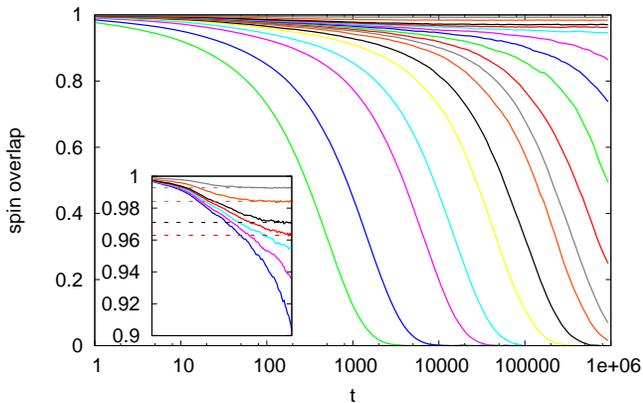}
    \vspace{0cm}
    \caption{Local spin-spin correlation function $C_s^{XOR}(t)$ for the XORSAT model at 
    different temperatures,
    from left to right $T$=0.7, 
0.65, 
0.6, 
0.58, 
0.56, 
0.55, 
0.54, 
0.535, 
0.53, 
0.525, 
0.52, 
0.515, 
0.51, 
0.505, 
0.5, 
0.48, 
0.45.
    The inset shows the comparison of the long time dynamics with the
   spin overlap that is obtained through the cavity method, which is reported as a dashed line for each temperature~\cite{xor_1,KZ10a}.
    }
    \label{spin-overlap}
  \end{center}
\end{figure}

In the formalism used above the local spin-spin correlation function for the 
XORSAT problem reads:
\beq\label{corr_s_XOR}
\begin{split}
&C^{XOR}_s(t) = \frac{1}{N} \sum_{i=1}^N \langle \sigma_i(t) \sigma_i(0) \rangle \ , \\
& \langle \sigma_i(t) \sigma_i(0) \rangle
=\langle GS_{XOR} | \hat{\sigma}_i^z e^{- \hat{H}_{XOR} t} \hat{\sigma}_i^z | GS_{XOR} \rangle  \ ,
\end{split}\eeq
where 
$\langle \us | GS_{XOR} \rangle$
is the ground state of $\hat{H}_{XOR}$ defined in Eq.~(\ref{GSXOR}). 
These correlation functions are shown in Fig.~\ref{spin-overlap}: for $T<T_d$,
they tend to a plateau $\overline{q}_s^{XOR}$ at long times. 
The dynamical correlation functions, here and in the following, 
are obtained using the Metropolis algorithm for a system of $N=60000$ spins. 
Sampling the initial configuration below $T_d$ is extremely hard with standard techniques, 
but it can be easily achieved
using the planting technique~\cite{MoSe,KZ09a,ZK10b}.

The plateau value $\overline{q}_s^{XOR}$ can be recovered from a statistical treatment
of the thermodynamical properties of Eq.~(\ref{Hpspin}) via the cavity method~\cite{xor_1,KZ10a}. 
In this treatment,
it is assumed that the system at long times decorrelates inside one of the typical ergodic components
$\a$ at temperature $T$; then
\beq
\overline{q}_s^{XOR} = \frac{1}N \sum_{i=1}^N \overline{ \langle \s_i \rangle_\a \langle \s_i \rangle_\a }
\eeq
where the overline indicates the equilibrium average over ergodic components.
The equivalence between these two quantities is shown in the inset of Fig.~\ref{spin-overlap}.

\begin{figure}[b]
  \begin{center}
    \includegraphics[width=\columnwidth]{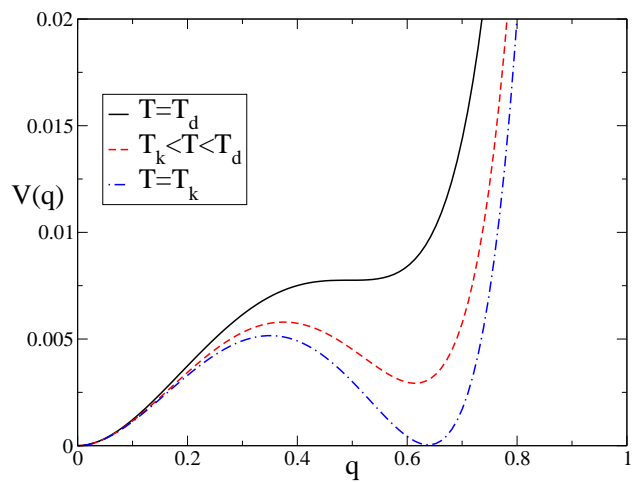}
    \vspace{0cm}
    \caption{
      A schematic representation of the Parisi-Franz potential~\cite{FP95,FP97} for a model
      in the RFOT class.
    }
    \label{fig:Vq}
  \end{center}
\end{figure}

A more detailed correspondence between statics and dynamics can be established by means of 
the Franz-Parisi thermodynamical potential~\cite{FP95,FP97}, defined as:
\beq\label{Vq_Ep}
\begin{split}
V_s^{XOR}&(q) = - \frac{T}{Z} \sum_{\underline{\sigma}} e^{-\beta E_{XOR}(\underline{\sigma})} \times \\
&  \times \log \Big[ \sum_{\underline{\sigma}'} e^{-\beta E_{XOR}(\underline{\sigma}')} \delta(q - q_s(\underline{\sigma},\underline{\sigma}') )\Big] \ ,
\end{split}\eeq
where $q_s(\underline{\sigma},\underline{\sigma}')  = \frac1N \sum_{i=1}^{N} \sigma_i \sigma_i'$
is the spin overlap function that measures the similarity of two configurations, $\sigma$ and $\sigma'$.
The potential $V_s^{XOR}(q)$ represents the free energy of a system of spins $\us'$ constrained to be
at fixed spin overlap $q$ from the spin system defined by $\us$. 
For mean-field models described by the RFOT theory, 
the potential $V_s^{XOR}(q)$ has the following properties,
see Fig.~\ref{fig:Vq}:
\begin{itemize}
\item It is a convex function with a single minimum at $q=0$ for $T>T_d$
\item For $T_d>T>T_k$ it displays a secondary minimum at $\overline{q}_s^{XOR}>0$, such that 
$V_s^{XOR}(\overline{q}_s^{XOR})>V_s^{XOR}(0)$. The difference between the two minima can be interpreted
as the free energy cost to force the two system to have a large overlap.
\item For $T<T_k$ the secondary minimum becomes thermodynamically stable, $V_s^{XOR}(\overline{q}_s^{XOR}) < V_s^{XOR}(0)$,
signaling the onset of replica symmetry breaking.
\end{itemize}
For the model under consideration, i.e. XORSAT with $p=c$, it turns out that 
$T_k=0$, as we already discussed; thus the third regime is not relevant
for the present discussion. Importantly enough, the value of the overlap $\overline{q}_s^{XOR}$ 
where the secondary minimum is attained when $T<T_d$ represents the plateau at which
the dynamical correlations of the spins converge in the long time limit.
Thus, the non-trivial
dynamical behavior can be understood in terms of thermodynamical quantities, 
which is the key feature of RFOT theory~\cite{FP95,FP97,FPRR11}.

\subsubsection{Plaquette overlap}

\begin{figure}
  \begin{center}
    \includegraphics[width=\columnwidth]{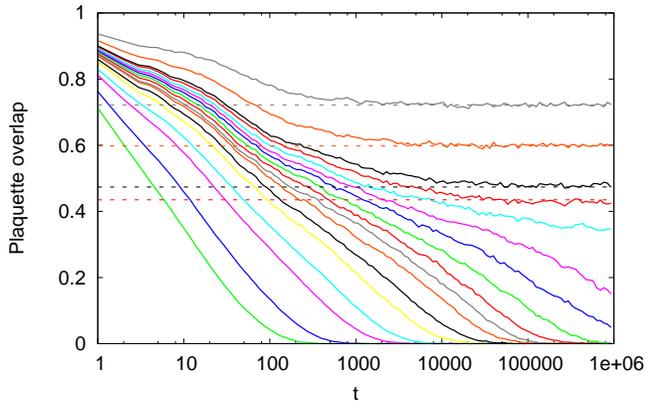}
    \vspace{0cm}
    \caption{Local plaquette-plaquette correlation functions $\phi_p^{XOR}(t)$ (solid lines),
    and the corresponding ``intra-state'' thermodynamical overlap (dashed lines) 
    computed with the cavity method~\cite{KZ10a}. Different curves represent the same temperatures as in Fig.~\ref{spin-overlap}.
}
    \label{Plaquette dynamics}
  \end{center}
\end{figure}

Let us now turn to the discussion of a second observable for the same model.
Instead of focusing on spin-spin correlations we study here the dynamics of 
a ``plaquette'' variable made of
three interacting spins, i.e. we consider:
\beq\label{corr_p}
\begin{split}
&C_p^{XOR}(t) = \frac{1}N \sum_{a=1}^N \langle\sigma_{i^1_a}(t)\sigma_{i^2_a}(t)\sigma_{i^3_a}(t)
\sigma_{i^1_a}(0)\sigma_{i^2_a}(0)\sigma_{i^3_a}(0) \rangle \ , \\ 
&\langle\sigma_{i^1_a}(t)\sigma_{i^2_a}(t)\sigma_{i^3_a}(t)
\sigma_{i^1_a}(0)\sigma_{i^2_a}(0)\sigma_{i^3_a}(0) \rangle 
= \\ & =  \langle GS_{XOR} | \hat{\sigma}_{i^1_a}^z\hat{\sigma}_{i^2_a}^z
\hat{\sigma}_{i^3_a}^z e^{- \hat{H}_{XOR} t} \hat{\sigma}_{i^1_a}^z
\hat{\sigma}_{i^2_a}^z\hat{\sigma}_{i^3_a}^z | GS_{XOR} \rangle  \ .
\end{split}\eeq
Because the average of a plaquette variable is 
$\la \sigma_{i^1_a} \sigma_{i^2_a} \sigma_{i^3_a} \ra= \tanh(\beta)$, it is better to consider
the connected and normalized function
\beq
\phi^{XOR}_p(t) = \frac{ C_p^{XOR}(t) - \tanh^2(\beta) }{ C_p^{XOR}(0) - \tanh^2(\beta) } \ .
\eeq
This correlation is shown in Fig.~\ref{Plaquette dynamics}
for different temperatures; it
displays a behavior slightly different from
the one shown in Fig.~\ref{spin-overlap}, but still below the dynamical
transition $T<T_d$ it does never relax and it reaches a plateau
whose value can be computed through the cavity method~\cite{KZ10a}, similarly to
what done in the previous case. This plateau value represents the 
overlap of plaquettes of spins between configurations belonging to the same state. 
This behavior is not surprising in light of the understanding of the dynamical
transition as the decomposition of the measure in many states, within the
RFOT theory picture. With such local dynamics, the system is not able
to escape from the initial state and the cavity method
allows us to access precisely this intra-state properties.

Because the dynamical behavior of the spin and plaquette overlap is qualitatively
equivalent, one would like to understand whether, from a thermodynamic 
point of view, the plaquette overlap can be used as an order parameter to constrain
the system in one metastable state, through the Franz-Parisi potential.
For the plaquette overlap, the Franz-Parisi potential $V_p^{XOR}(q)$ 
is defined as in~(\ref{Vq_Ep}), with $q_s(\us,\us')$ replaced by
\beq
q_p(\us,\us') = \frac1N \sum_{a=1}^N \sigma_{i^1_a} \sigma_{i^2_a} 
\sigma_{i^3_a} \sigma_{i^1_a}' \sigma_{i^2_a}' \sigma_{i^3_a}' \ .
\eeq
An explicit calculation of this quantity is possible, by 
exploiting the mapping towards the
non-interacting model discussed in Section~\ref{sec:mapping}, thanks to the crucial
fact that the plaquette overlap in XORSAT becomes the spin overlap in the KCM.
Therefore $V_p^{XOR}(q) = V_s^{KCM}(q)$. The computation is done in Appendix~\ref{AppendixVq}
and it shows explicitly that the function $V_p^{XOR}(q)$ is a convex function at all temperatures, 
with a single minimum at 
$q = \tanh^2(\beta)$ corresponding to the thermodynamic value of $q_p$ in the paramagnetic state.
We conclude that $V_p^{XOR}(q)$ does not signal the presence of a metastable state and not even 
provide an interpretation for the change in the dynamical behavior 
of the plaquette correlation in Eq.~(\ref{corr_p}) when $T<T_d$.

This example shows that in order to find a satisfactory thermodynamic description
that accounts for the dynamical behavior 
it is important to select the correct order parameter which in 
the case of the XORSAT model is the spin overlap.
Even if, a priori, imposing a given spin overlap between two configurations
might not seem very different from fixing the plaquette overlap, 
this leads to a qualitative different behavior of the corresponding Franz-Parisi potential.
The reason is readily found by thinking to the structure of the energy landscape of XORSAT, 
and in particular, as in the case under study, when there is a single ground state.
Up to an overall constant the ground state has energy zero and above it there are $N$ excited 
configurations that violate exactly one interaction (i.e. one plaquette has value $-1$). 
These configurations are at extensive Hamming distance 
among each other and with the ground state, therefore their spin overlap with the ground state is very close
to zero.
However, in terms of plaquettes,
they are at distance 1 from the ground state 
and at distance 2 among each other, therefore their plaquette overlap with
the ground state and among each other is very large.
This implies that the requirement to fix a high value of the plaquette overlap in the Franz-Parisi potential
does not select a state and on the contrary 
allows to sum over configurations belonging to different states.
For this reason the overlap $q_p(\us,\us')$ does not work as an order parameter
that signals the emergence of many metastable states that trap the dynamics. This was also
recognized in the context of finite-dimensional plaquette models~\cite{JG05}.

The situation is somehow analogous to what happens if one considers a ferromagnetic Ising
model and adds an infinitesimal field which couples to
$s_i s_j$. Clearly, this field does not break the symmetry and is therefore unable to detect the phase transition by constraining
the system in one of the two symmetry-breaking pure states.
However, the dynamical correlations of $s_i s_j$
would clearly show a plateau in the low temperature phase if the system is constrained
into one of its pure states.

\subsection{Kinetically constrained model}

We turn now to the discussion of the kinetically constrained model,
introduced in Section~\ref{Sec:KCM}.
Due to the mapping onto XORSAT, non-trivial properties in the dynamics
of the KCM are expected to appear  
in spite of the simplicity of the energy function Eq.~(\ref{energy-h}).

\begin{figure}
  \begin{center}
    \includegraphics[width=\columnwidth]{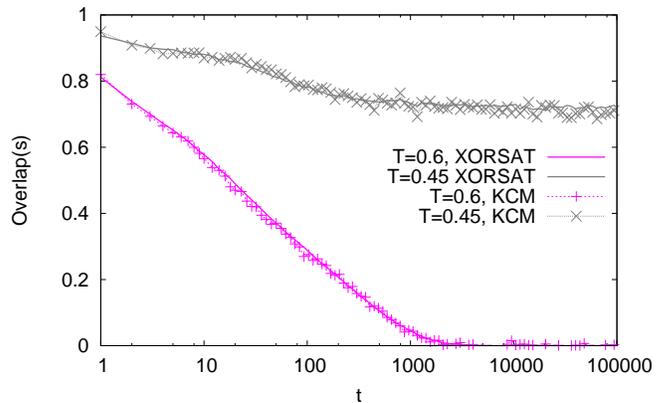}
    \vspace{0cm}
    \caption{Comparison between the (connected and normalized) 
      spin-spin correlation function of the KCM, $\phi^{KCM}_s(t)$ (points),
      with the plaquette-plaquette correlation $\phi^{XOR}_p(t)$ of XORSAT (solid lines). 
      The grey and the pink curves refer
      to diffent temperatures, respectively $T=0.6$ and $T=0.45$.}
    \label{Comparison}
  \end{center}
\end{figure}

The equilibrium local spin-spin correlations for this model read:
\beq\label{corr_h}
\begin{split}
& C_s^{KCM}(t) = \frac1N \sum_{a=1}^N \langle s_a(t) s_a(0) \rangle \ , \\
&\langle s_a(t) s_a(0) \rangle  
=\langle GS_{KCM} | \hat{s}^x_a \, e^{- \hat{H}_{KCM} t} \, \hat{s}^x_a | GS_{KCM} \rangle \ , \\
&\phi^{KCM}_s(t) = \frac{ C_s^{KCM}(t) - \tanh^2(\beta) }{ C_s^{KCM}(0) - \tanh^2(\beta) } \ , \\
\end{split}\eeq
where 
$\langle \underline{s} | GS_{KCM} \rangle$
is the ground state of $\hat{H}_{KCM}$ defined in Eq.~(\ref{GSKCM}). 
Thanks to the duality transformation the ground state of $\hat{H}_{KCM}$ is mapped into that
of $\hat{H}_{XOR}$ and the time evolution operator as well. Moreover the observable
$\hat{s}^x_a$ is mapped in $\hat{\sigma}^z_{i^1_a}\hat{\sigma}^z_{i^2_a}\hat{\sigma}^z_{i^3_a}$
in the XORSAT problem.
Thus, the spin correlation in the KCM, Eq.~(\ref{corr_h}) is 
equivalent to the plaquette correlation
in the XORSAT problem, Eq.~(\ref{corr_p}): $C^{KCM}_s(t) = C_p^{XOR}(t)$.
This equivalence is confirmed numerically in Fig.~\ref{Comparison}
at two different temperatures, $T=0.45$ and $T=0.6$. 
The correlation functions of the KCM have been obtained using
Metropolis dynamics in discrete time for a system with $N=60000$ spins.
In the case of the KCM, the problem of generating the initial equilibrium
configuration is of course completely trivial.

Thus, the dynamics of the spins in the KCM 
is the same shown in Fig.~\ref{Plaquette dynamics} for the plaquettes in XORSAT. 
It reveals non-trivial features 
and in particular a dynamical transition 
at the same critical temperature $T_d=0.5098(2)$.
Such a transition cannot be understood in terms of 
a simple thermodynamical reasoning on
the basis of the Franz-Parisi potential for the spin overlap of the KCM.
Indeed, as we already discussed, $V_s^{KCM}(q) = V_p^{XOR}(q)$ and both are
trivial functions of $q$, see Appendix~\ref{AppendixVq}.
Indeed, due to the non-interacting nature of the spins of the KCM,
from the static point of view the overlap 
$q_s(\underline{s},\underline{s}') = \frac1N \sum_{a=1}^N s_a s_a'$ 
is not expected to 
select any state and it cannot work as a static order parameter.
This phenomenology is very reminiscent of other kinetically constrained models.
At first sight thus, in the KCM one is not able
to understand the dynamic transition as the consequence of a complex thermodynamic
behavior.

However, the mapping in Eq.~(\ref{inversemapping})
teaches us that {\it there is a static order parameter
that does this work}, i.e. what corresponds to the overlap between spin configurations
of the XORSAT model. 
Expressed in terms of the spins $s_a$, it becomes:
\beq
q_{nl}(\underline{s},\underline{s}') = \frac{1}{N}\sum_{i=1}^N \prod_{a=1}^N (s_a s_a')^{y^a_i} \ .
\eeq
However this quantity is a highly non local observable and then it cannot be easily and
naturally recognized, especially if the mapping is not explicitly given
(note that in particular the $y^a_i$ depend on the particular instance of random XORSAT under
consideration).

\section{Discussion}
\label{sec:discussion}

\subsection{Summary of the results \\ for the Bethe lattice model}

In this work we have shown that it is possible to establish a
mapping between the equilibrium dynamics of a 
glassy mean-field model described by the Random First-Order Transition 
theory, i.e. the XORSAT model, and a Kinetically Constrained Model.

We have discussed the role of the overlap function that is chosen
in the framework of RFOT theory to work as order parameter and signal 
the presence of metastable states. Overlap functions that couple
different observables define different distances between configurations
and thus might fail to select a particular state. This is the case 
of the plaquette overlap in the XORSAT model. 
This observable is directly mapped into the spin overlap in the KCM. 
Therefore, the thermodynamic properties, and in particular
the Franz-Parisi potential, are trivial when the plaquette overlap is considered
in XORSAT, or the spin overlap is considered in the KCM. 

However the dynamics of all local correlations shows no relaxation
below a critical temperature $T_d$. Within the XORSAT model, this is signaled by a static
computation of the Franz-Parisi potential for the spin overlap, as usual in RFOT theory.
For the KCM, thanks to the mapping back to the XORSAT model,
one can recast the non-trivial dynamical behavior 
within a thermodynamical description whose basic degrees of freedom 
however are hidden non-local functions of the spins.

Thus, this example illustrates that --at least in this case-- one is able to resolve the apparent 
dichotomy that exist between static and dynamic properties 
tracing them back to the physics of RFOT theory. This is not obvious from the point of view of
the KCM: if one were to study it without any knowledge on the mapping on the RFOT model, identifying
the correct order parameter would be impossible.

\subsection{Finite dimensional models}

As we discussed in the introduction, the mapping we used for the Bethe lattice model
was already used extensively 
in a class of finite dimensional plaquette models~\cite{Ga02,JBG05,JG05,CB12}.
However, these plaquette models do not seem to be described by the RFOT theory,
because the energy cost due to the presence of an interface between two
low energy states does not grow with system size~\cite{JG05,JBG05,CB12,JB12}.

On the other hand, we have shown in section~\ref{sec:Gmapping} 
that the mapping between the $p$-spin model and
the KCM holds quite generally, under the only requirements that {\it i)}~each 
spin is involved in exactly $p$ interactions, {\it ii)}~each interaction involves 
exactly $p$ spin, and {\it iii)}~the ground state is unique.
Therefore, it might be possible to devise a finite-dimensional model
(possibly in three or higher dimensions)
that satisfies these requirements and at the same time is well described by the RFOT scenario.
This could be possibly achieved by extensions of the so-called gonihedric models~\cite{LJ00,DEJP02}
or of the three dimensional plaquette model considered in~\cite{KZ11}. A numerical computation
of the Franz-Parisi potential for the spin overlap in these models, following the strategy of~\cite{FMY09,CCGGGPV10},
could be instructive.

Such a model could be quite useful to investigate the relation between the
RFOT and KCM pictures in a finite dimensional setting. We recall, however, that
the mapping implies that $T_k=0$, because the free energy must be analytic at all
temperatures, therefore a RFOT scenario with a positive Kauzmann temperature
would not fit in this analysis.

Whether a similar mapping can be constructed in more general models remains an open 
problem~\cite{KPS97}.
Finally, it would be very interesting to check whether a KCM with a continuous 
transition~\cite{SDCA10} could be constructed, for which a similar mapping exists.

\acknowledgments

We thank L.~Berthier, G.~Biroli, J.~P.~Garrahan,
D.~Gosset, R.~L.~Jack, C.~Laumann, V.~Martin-Mayor,
M.~Sellitto and G.~Semerjian for many important discussion on this work.

\bibliography{HS,SAT}

\bibliographystyle{mioaps}

\begin{widetext}

\appendix

\section{Franz-Parisi potential for independent spins}\label{AppendixVq}

Here we report the calculation of the Franz-Parisi potential for the plaquette overlap in XORSAT,
$V^{XOR}_p(q)$, that is mapped onto the same potential for the spin overlap in the KCM, $V^{KCM}_s(q)$.
Since in the KCM the spins are independent, the latter is trivially computed. We have:
\beq
V^{XOR}_p(q) = V^{KCM}_s(q) = - \frac{1}{\beta Z} \sum_{\underline{\sigma}} e^{-\beta E_{KCM}(\underline{\sigma})} \log \Big[ \sum_{\underline{\sigma}'} e^{-\beta E_{KCM}(\underline{\sigma}')} \delta(q - q_s(\underline{\sigma},\underline{\sigma}') \Big]
\eeq
where the spin overlap is
$q_s(\underline{\sigma},\underline{\sigma}') = \frac1N \sum_i \sigma_i \sigma'_i$
for a system of free spins in magnetic field
$E_{KCM}(\underline{\sigma}) = - \sum_i \sigma_i$.
Exploiting the relation 
$\log x = \lim_{n\to 0} \partial_n x^n$ 
 the Franz-Parisi potential becomes:
\beq
 \begin{array}{l}
\displaystyle V^{KCM}_s(q) = - \lim_{n\to 0} \partial_n  \sum_{\underline{\tau}} 
\frac{e^{\beta \sum_i \tau_i}}{\beta (2 \cosh \beta)^N} \sum_{\underline{\tau}^1,\dots,\underline{\tau}^n }
e^{\beta \sum_{i,a} \tau_i^a} \, \prod_{a=1}^n\delta\Big( N q - \sum_i \tau_i \tau_i^a \Big)

  \\ \vspace{-0.2cm} \\

\displaystyle = - \lim_{n\to 0} \partial_n 
\frac{1}{\beta (2 \cosh \beta)^N}  \int {\rm d}\lambda_1 \dots {\rm d}\lambda_n \, e^{ - \beta N q \sum_{a=1}^n \lambda_a}
 \sum_{\underline{\tau}, \underline{\tau}^1,\dots,\underline{\tau}^n }
e^{\beta (\sum_{i} \tau_i + \sum_{i,a} \tau_i^a) + \beta \sum_{i,a} \lambda_a \tau_i \tau_i^a}

\\ \vspace{-0.2cm} \\

\displaystyle = - \lim_{n\to 0} \partial_n 
\frac{1}{\beta (2 \cosh \beta)^N}  \int {\rm d}\lambda_1 \dots {\rm d}\lambda_n \, e^{- \beta N q \sum_{a=1}^n \lambda_a}
\Big(  \sum_{\tau, \tau^1,\dots,\tau^n }
e^{\beta (\tau + \sum_{a} \tau^a) + \beta \sum_{a} \lambda_a \tau \tau^a} \Big)^N

\\ \vspace{-0.2cm} \\

\displaystyle = - \lim_{n\to 0} \partial_n
\frac{1}{\beta (2 \cosh \beta)^N}  \int {\rm d}\lambda \, e^{- \beta N q n \lambda + N \log f(\lambda)}

 \end{array}
\eeq
with
\beq
 \begin{array}{l}
f(\lambda) =  \sum_{\tau, \tau^1,\dots,\tau^n }
e^{\beta (\tau + \sum_{a} \tau^a) + \beta \lambda \sum_{a} \tau \tau^a}

= \sum_{\tau}
e^{\beta \tau}\Big[ 2 \cosh\Big( \beta ( 1 + \lambda \tau )  \Big) \Big]^n
\\ \vspace{-0.2cm} \\
= e^{\beta}\Big[2 \cosh\Big( \beta ( 1 + \lambda ) \Big) \Big]^n
+ e^{-\beta}\Big[ 2\cosh\Big( \beta ( 1 - \lambda ) \Big) \Big]^n

 \end{array}
\eeq

Developing for small $n$
\beq
\displaystyle f(\lambda) = 2 \cosh(\beta) \Big[ 1 + \frac{n}{2\cosh\beta} (
e^{- \beta}\log\Big[2 \cosh\Big( \beta ( 1 - \lambda ) \Big) \Big]
+ e^{\beta}\log\Big[ 2\cosh\Big( \beta ( 1 + \lambda ) \Big) \Big]) \Big]
\eeq

Thus
\beq\label{Vq-noninterspin}
 \begin{array}{l}
\displaystyle  V^{KCM}_s(q) = - \frac{1}{\beta} \lim_{n\to 0} \partial_n  
 \int {\rm d}\lambda \, e^{- \beta N q n \lambda + \frac{N n}{2\cosh\beta}  \Big( e^{- \beta}\log \Big[ 2 \cosh\Big( \beta ( 1 - \lambda ) \Big) \Big]
+ e^{\beta}\log \Big[ 2 \cosh\Big( \beta ( 1 + \lambda)  \Big) \Big]  \Big)}

\\ \vspace{-0.2cm} \\

\displaystyle \simeq \frac{N}{\beta} \min_{\lambda} \Big[ \beta q \lambda - \frac{1}{2\cosh\beta} \Big( e^{- \beta}\log \Big[ 2 \cosh\Big( \beta ( 1 - \lambda )  \Big) \Big]
+ e^{\beta}\log \Big[ 2 \cosh\Big( \beta ( 1 + \lambda ) \Big) \Big]  \Big)  \Big]
 \end{array}
\eeq

and
\beq
q(\lambda) = \frac{1}{2\cosh \beta} \Big( e^{\beta} \tanh(\beta(1+\lambda)) - e^{-\beta} \tanh(\beta(1-\lambda)) \Big)
\eeq
The function~(\ref{Vq-noninterspin}) is a convex function at all temperatures,
with a single minimum at $q^{\ast}=q(\lambda=0)=(\tanh{\beta})^2$, such that 
$V_s^{KCM}(q^{\ast})=F(T)= - \frac{N}{\beta}\log 2\cosh(\beta)$.

\end{widetext}


\end{document}